%% file: ms.tex
\documentclass[fleqn,usenatbib]{mnras}

\usepackage{newtxtext,newtxmath}
\DeclareRobustCommand{\VAN}[3]{#2}
\let\VANthebibliography\thebibliography
\def\thebibliography{\DeclareRobustCommand{\VAN}[3]{##3}\VANthebibliography}
\usepackage{graphicx}	% Including figure files
\usepackage{amsmath}	% Advanced maths commands
\usepackage{orcidlink}
\usepackage{pgfplots}

\title{High resolution radio observations of the Chamaeleon star-forming region}

\author[E. García Valencia et al.]{Ernesto García Valencia,$^{1}$
Laurent Loinard,$^{2,3,4}$\orcidlink{0000-0002-5635-3345}
Arnaud Belloche,$^{5}$\orcidlink{0000-0003-0046-6217}
Gisela N. Ortiz León,$^6$\orcidlink{0000-0002-2863-676X}\newauthor
Sergio A.\ Dzib,$^5$\orcidlink{0000-0001-6010-6200}
Cherie K.\ Day,\orcidlink{0000-0002-8101-3027}
Roopesh Ojha,$^7$\orcidlink{0000-0003-2556-8623}\\
\\
$^1$ Departamento de Investigación en Física, Universidad de Sonora, Apartado Postal 83190, Hermosillo, Sonora, México\\
$^{2}$ Instituto de Radioastronom\'ia y Astrof\'isica, Universidad Nacional Aut\'onoma de M\'exico, Morelia 58341, M\'exico\\
$^{3}$ Black Hole Initiative at Harvard University, 20 Garden Street, Cambridge, MA 02138, USA \\
$^{4}$ David Rockefeller Center for Latin American Studies, Harvard University, 1730 Cambridge Street, Cambridge, MA 02138, USA\\
$^5$ Max-Planck-Institut für Radioastronomie, Auf dem Hügel 69, D-53121 Bonn, Germany\\
$^6$ Instituto Nacional de Astrofísica, Óptica y Electrónica, Apartado Postal 51 y 216, 72000 Puebla, México\\
$^7$ NASA Headquarters, 300 E St SW, Washington, DC 20546, USA
}

\date{Accepted \today; Received \today; in original form \today}

\pubyear{\the\year{2025}}

\begin{document}
\label{firstpage}
\pagerange{\pageref{firstpage}--\pageref{lastpage}}
\maketitle

\begin{abstract}
We report on large-scale radio observations of the Chamaeleon star-forming region obtained with the Australia Telescope Compact Array (ATCA) that led to the definite detection of five young stars and the tentative detection of five more. As in other regions surveyed in the radio domain, the majority of detected sources are fairly evolved low-mass T Tauri stars, but we also detect one protostellar object (Ced\,110~IRS4) and one Herbig Ae/Be star. With the exception of the protostellar source, the radio emission mechanism is likely of non-thermal origin. The three brightest radio stars identified with ATCA were subsequently observed with the Australian Long Baseline Array (LBA) and one, J11061540$-$7721567 (Ced\,110~IRS2), was detected at three epochs. This confirms the non-thermal nature of the radio emission in that specific case, and enabled accurate radio position measurements. Comparison with predictions from Gaia DR3 strongly suggests that this star is a binary system with an orbital period of order 40 years; additional LBA observations in the next decades would enable accurate determinations of the individual stellar masses in that system.    
\end{abstract}

\begin{keywords}
astrometry --- stars: activity --- stars: formation --- stars: pre-main-sequence --- techniques: interferometric
\end{keywords}

%%%%%%%%%%%%%%%%%%%%%%%%%%%%%%%%%%%%%%%%%%%%%%%%%%

%%%%%%%%%%%%%%%%% BODY OF PAPER %%%%%%%%%%%%%%%%%%

\section{Introduction} \label{sec:intro}

\begin{figure*}
    \centering
    \includegraphics[width=8.5 cm, angle=0]{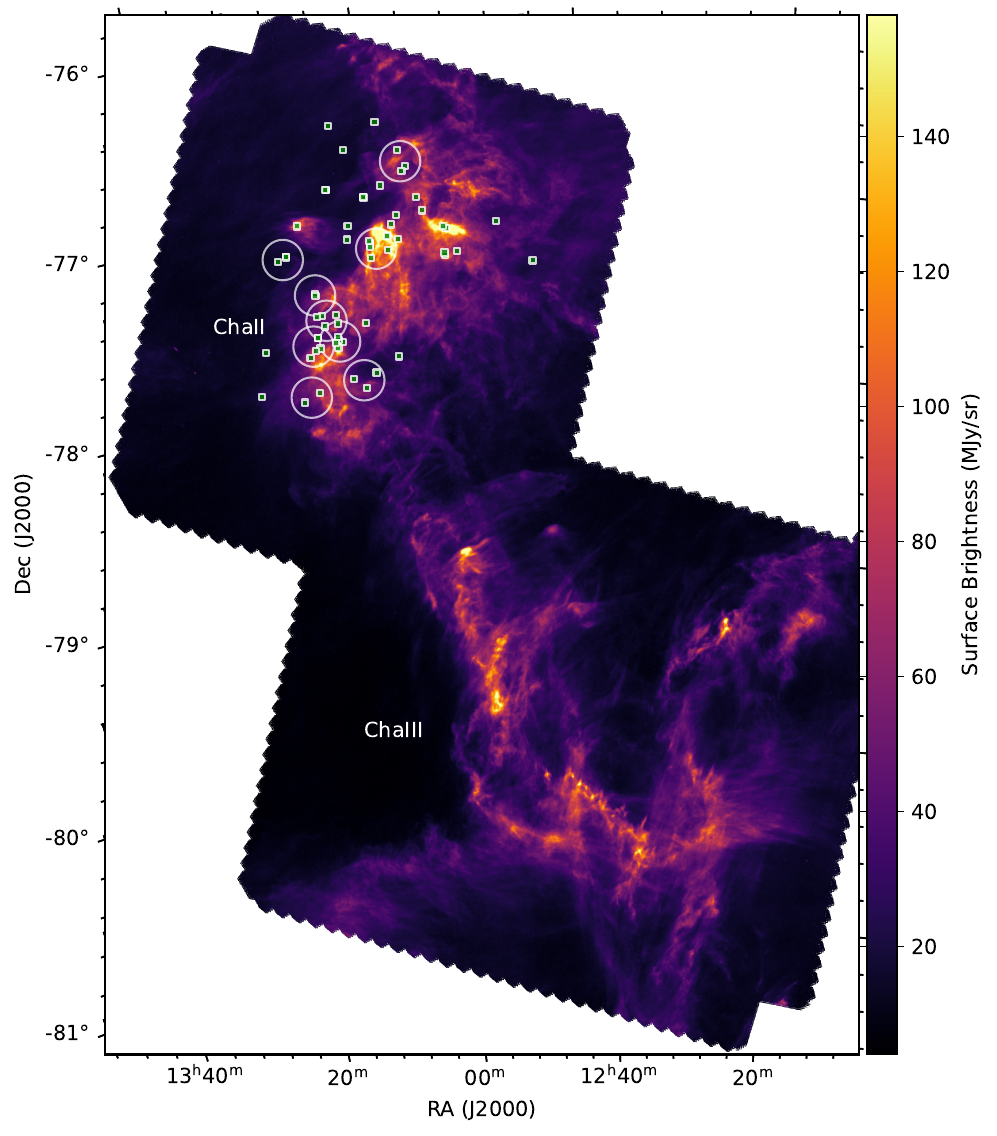}
    \includegraphics[width=8.5 cm, angle=0]{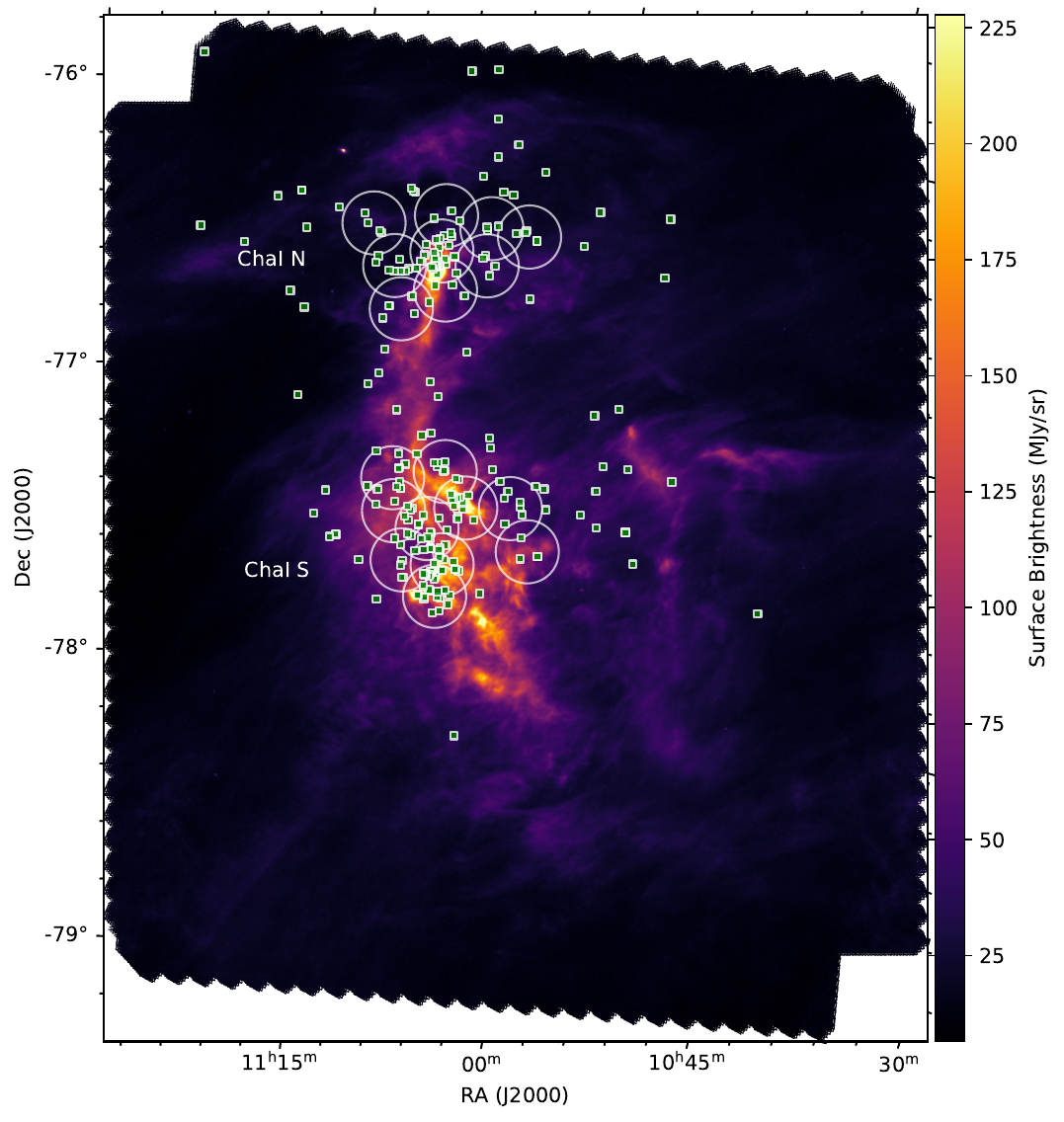}
    \caption{Herschel 250 $\mu$m maps of the Cha\,II/III regions (left) and the Cha\,I sub-regions (right) with the observed ATCA fields overlaids as circles (with a diameter of 9 arcmin corresponding to the field of view of ATCA at 5.5 GHz). Squares indicate the positions of known young stars in each field (from \citealt{Alcala+2008} in Cha\,II and \citealt{Esplin+2017} in the case of Cha\,I).}
    \label{fig:herschel}
\end{figure*}

The Chamaeleon cloud complex is a prominent star-forming region in the southern hemisphere centered roughly around RA= 12$^h$, Dec = $-78^\circ$ and comprising three main dark clouds \citep[Cha\,I, Cha\,II, and Cha\,III; ][see Figure \ref{fig:herschel}]{Schwartz1977}, each a few square degrees in size. The large-scale structure of the clouds has been studied using extinction maps based on star counts \citep[e.g.,][]{Dobashi+2005}, molecular mapping \citep{Mizuno+2001}, and far-infrared imaging \citep{Andre+2010}. Overall, the gas mass, the extinction, and the density of young stars are low in Chamaeleon compared to many other star-forming regions. In addition, Chamaeleon is an isolated complex where contamination by other star-forming regions is minimal. This combination of properties can facilitate detailed studies, so Chamaeleon has been a popular target for research on low-mass star formation over the last several decades.

The Cha\,I cloud, further subdivided into Cha\,I\,N (north) and Cha\,I\,S (south) (Figure \ref{fig:herschel}), is the richest in the complex with a total population of about 250 pre-main sequence stars in the recent survey by \citet{Esplin+2017}. In contrast, Cha\,II contains less than 100 stellar members \citep{Alcala+2008,Spezzi+2008}. According to the most recent studies, Cha\,I and Cha\,II have a similar age of order $2.0\pm0.5$ Myr \citep{Galli+2021}. Finally, the Cha\,III cloud appears to be devoid of on-going star-formation activity \citep{Luhman2008,Belloche+2011}; this may indicate that it represents an earlier evolutionary phase where star-formation is yet to occur.

The issue of the distances to the various clouds in Chamaeleon has been extensively discussed ever since its identification as a nearby star-forming region. Early estimates for Cha\,I varied between about 115 and 215 pc \citep[][and references therein]{Grasdalen+1975,Hyland1982,Schwartz1992}, whereas values as large as 400 pc were reported for Cha\,II \citep{Graham+1988}. Combining several methods, \citet{Whittet+1997} obtained 160$\pm$15 pc and 178$\pm$18 pc for Cha\,I and Cha\,II, respectively, while \citet{Bertout+1999} obtained 168$\pm$14 pc from a scant handful of Hipparcos parallaxes in Cha\,I. From Gaia DR1 data \citep{GaiaDR1}, \citet{Voirin+2018} obtained 179$\pm$22 pc, 181$\pm$17 pc, and  199$\pm$20 pc for Cha\,I, Cha\,II and Cha\,III, respectively. Using more recent Gaia DR2 results \citep{GaiaDR2}, \citet{Rocca+2018} found different distances from the two sub-components of Cha\,I: 192.7$\pm$0.4 pc and 186.5$\pm$0.7 pc for Cha\,I\,N and Cha\,I\,S, respectively. Nearly simultaneously, \citet{dzib2018} also used Gaia DR2 data and determined mean parallaxes of $5.21\pm0.01$ mas and $5.05\pm0.02$ mas, corresponding to distances of $191.9\pm0.4$\,pc and $198.0\pm0.8$\,pc for Cha\,I and Cha\,II, respectively.  These values were confirmed by a reanalysis of the Gaia DR2 results by \citet{Galli+2021} who reported 191.4$\pm$0.8 pc for Cha\,I\,N 186.7$\pm$1.0 pc for Cha\,I\,S, and 197.5$\pm$1.0 pc for Cha\,II.

In comparison with other wavelength ranges, the radio domain has been poorly explored in Chamaeleon. In his review, \citet{Luhman2008} does not mention any radio observation at all. \citet{Lehtinen+2003} observed a single-field in the Cederblad\,110 region in Cha\,I\,S at $\lambda$ = 3 and 6 cm, and reported the detection of three young stellar objects -- in addition to a handful of background sources. \citet{Brown+1996,Brown+2004} report on the radio detection with the Australia Telescope Compact Array (ATCA) of six\footnote{They mention seven detections, but only explicitly report six.} young stars in Chamaeleon, including two of those also reported by \citet{Lehtinen+2003}. Yet, one would expect to find radio bright young stars in Chamaeleon, because extensive X-ray observations with ROSAT \citep{Feigelson+1993,Alcala+1995,Alcala+2000}, XMM-Newton \citep{Stelzer+2004,Robrade+2007} and Chandra \citep{Feigelson+2004} have revealed a nourished population of active young stars in the region. Given the known relation between X-ray and radio fluxes \citep{Gudel1993}, radio-bright stellar sources ought to exist in the Chamaeleon clouds. 

In this paper, we report on an extensive search for compact radio sources in Cha\,I and Cha\,II performed with ATCA. This led to the definite detection of five young stars and the tentative detection of five more. For the three brightest and most promising radio young stars detected with ATCA, we have obtained complementary Long Baseline Array (LBA) observations that will also be reported here. The observations will be described in Section \ref{sec:obs}, the results will be presented in Section \ref{sec:results} and discussed in Section \ref{sec:discus}. Section \ref{sec:conclusions} contains our conclusions and some perspectives.

\input{table1}

\section{Observations} \label{sec:obs}

\subsection{ATCA observations} \label{sec:atca}

The distribution of young stars identified in Chamaeleon by \cite{Esplin+2017} is shown, overlaid on a 250 $\mu$m Herschel image, in Figure \ref{fig:herschel}. It occupies more than ten square degrees on the sky, and it would be very time consuming to survey systematically this entire area. Instead, we decided to map only the sub-regions with the highest concentrations of young stars. A judicious choice of 28 pointing centers (nine pointings in Cha\,IN, ten in Cha\,IS, and nine in Cha\,II; see Figure \ref{fig:herschel}) enabled us to include two thirds of the young stars in Cha\,I and 55\% of those in Cha\,II. The center positions of these fields are provided in Table \ref{tab:ATCA_obslog}.

The observations (project code C2682) were obtained on 2 March 2013 with the Australia Telescope Compact Array (ATCA; \citealt{Frater+1992}) located near Narrabri in New South Wales (Australia). The array was in its most extended (6A) configuration that offers east-west baselines up to 6 km long. Two frequency bands (centered at 5.5 and 8 GHz) and two linear polarizations were observed simultaneously. Each frequency band was 2 GHz wide and covered by 2048 channels each 1 MHz wide with the Australia Telescope Compact Array Broad-band Backend (CABB) correlator \citep{CABB2011}. Observations of the 28 Chamaeleon fields were distributed throughout the 9 hours of observation to optimize the ({\it u,v}) coverage. Each field was visited three or four times for a total on-source integration of 13 to 16 minutes.

The data were calibrated following standard procedures using the Miriad software package \citep{MIRIAD1995}. The complex gains were measured using observations of the nearby quasar 1057--797 obtained roughly every 25 minutes throughout the observing run. The quasar 1934--638 served as bandpass and flux calibrator. After calibration, the visibilities were imaged in CASA \citep{CASA2022} using a natural weighting scheme (Briggs robust parameter set to 2; \citealt{Briggs1995}). At both frequencies, we produced images with $4,096\times4,096$ pixels each 0.2 arcsec. Pixels outside of a primary beam response of 0.2 (corresponding to a distance from the field center of 6.6 arcmin and 4.7 arcmin at 5.5 GHz and 8.0 GHz, respectively) were blanked. The resulting synthesized beams and noise levels are provided in Table \ref{tab:ATCA_obslog}.

\begin{figure*}
    \centering
    \includegraphics[width=17cm]{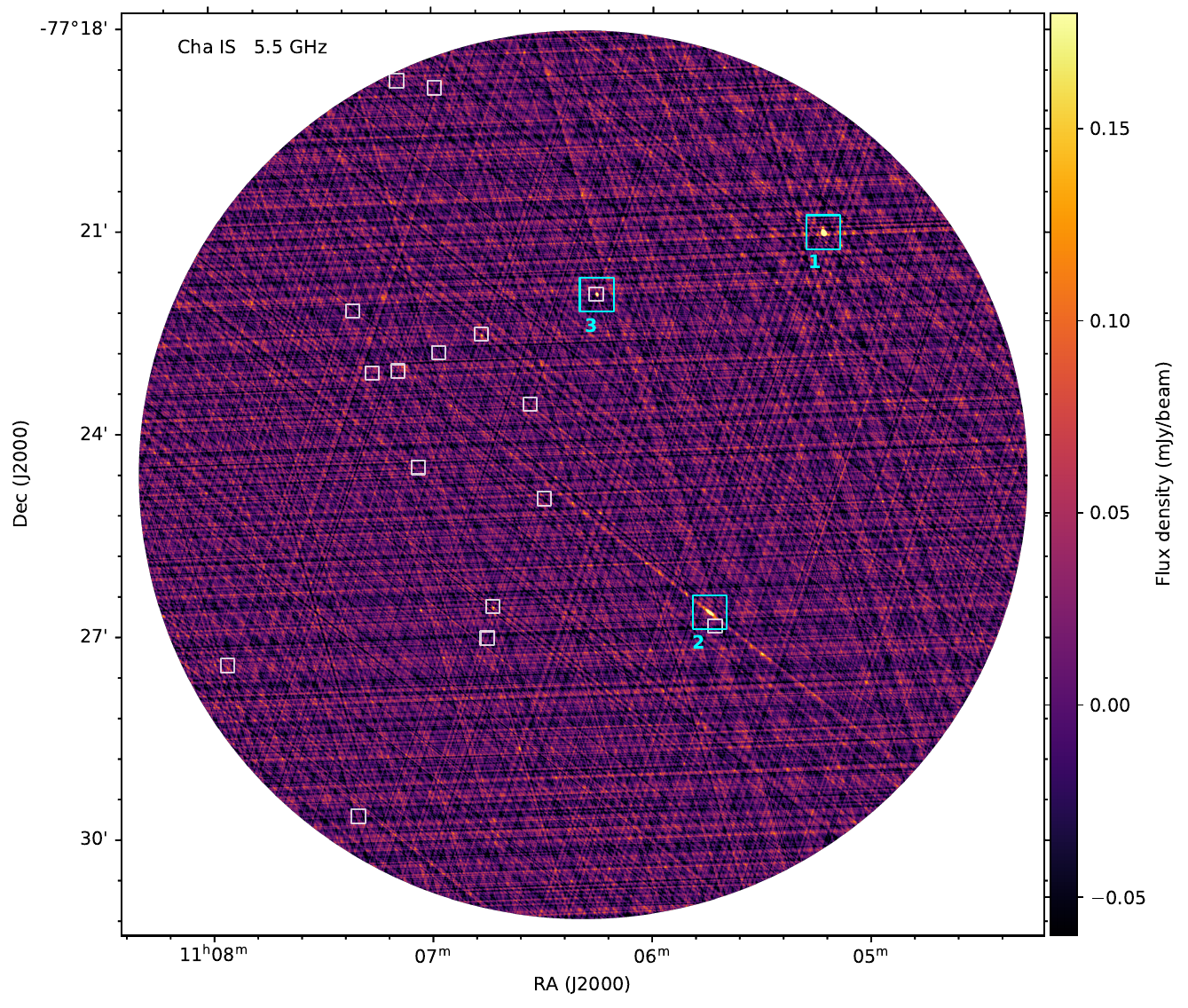}
    \includegraphics[width=4cm]{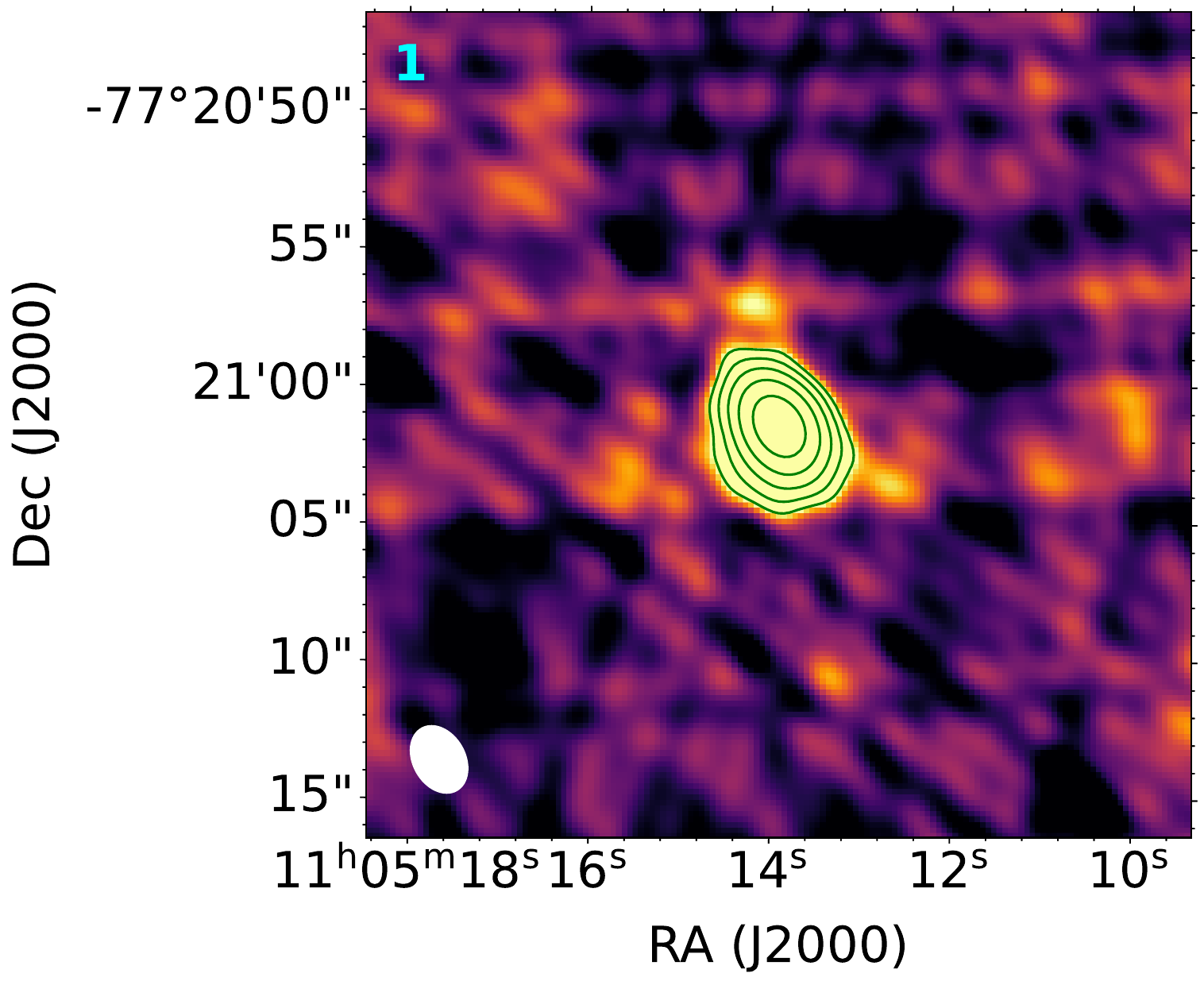}
    \includegraphics[width=4.1cm]{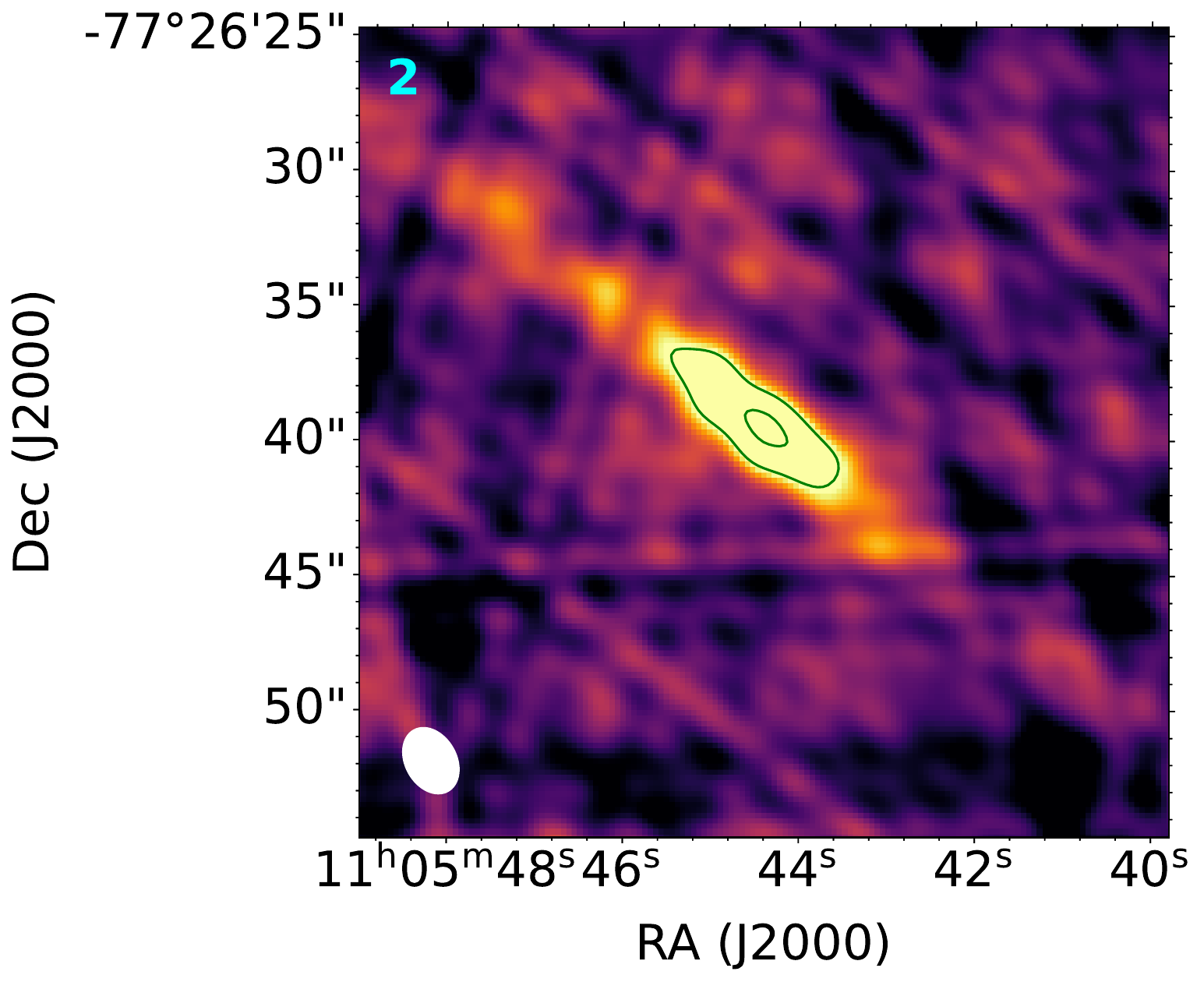}
    \includegraphics[width=4cm]{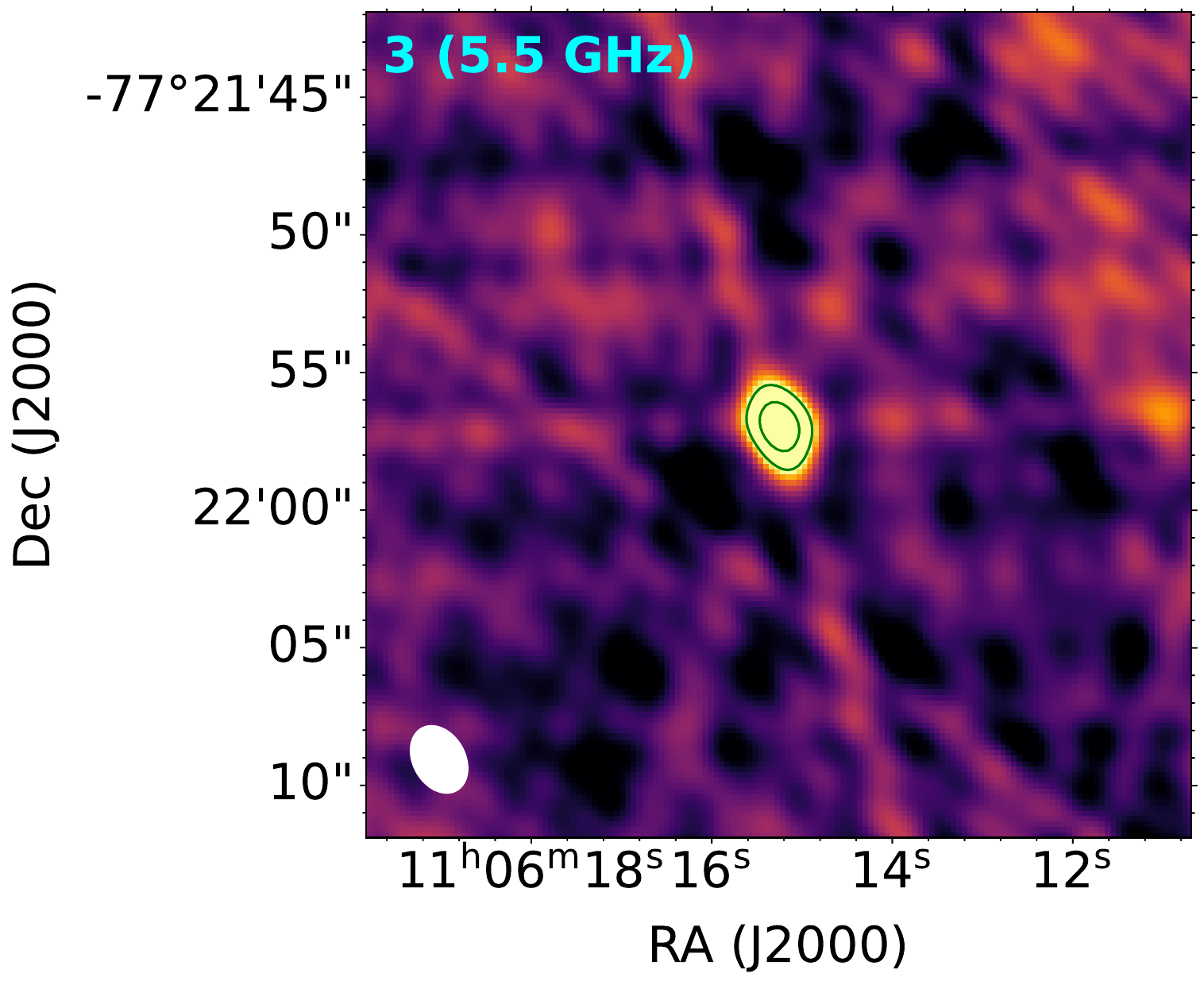}
    \includegraphics[width=4cm]{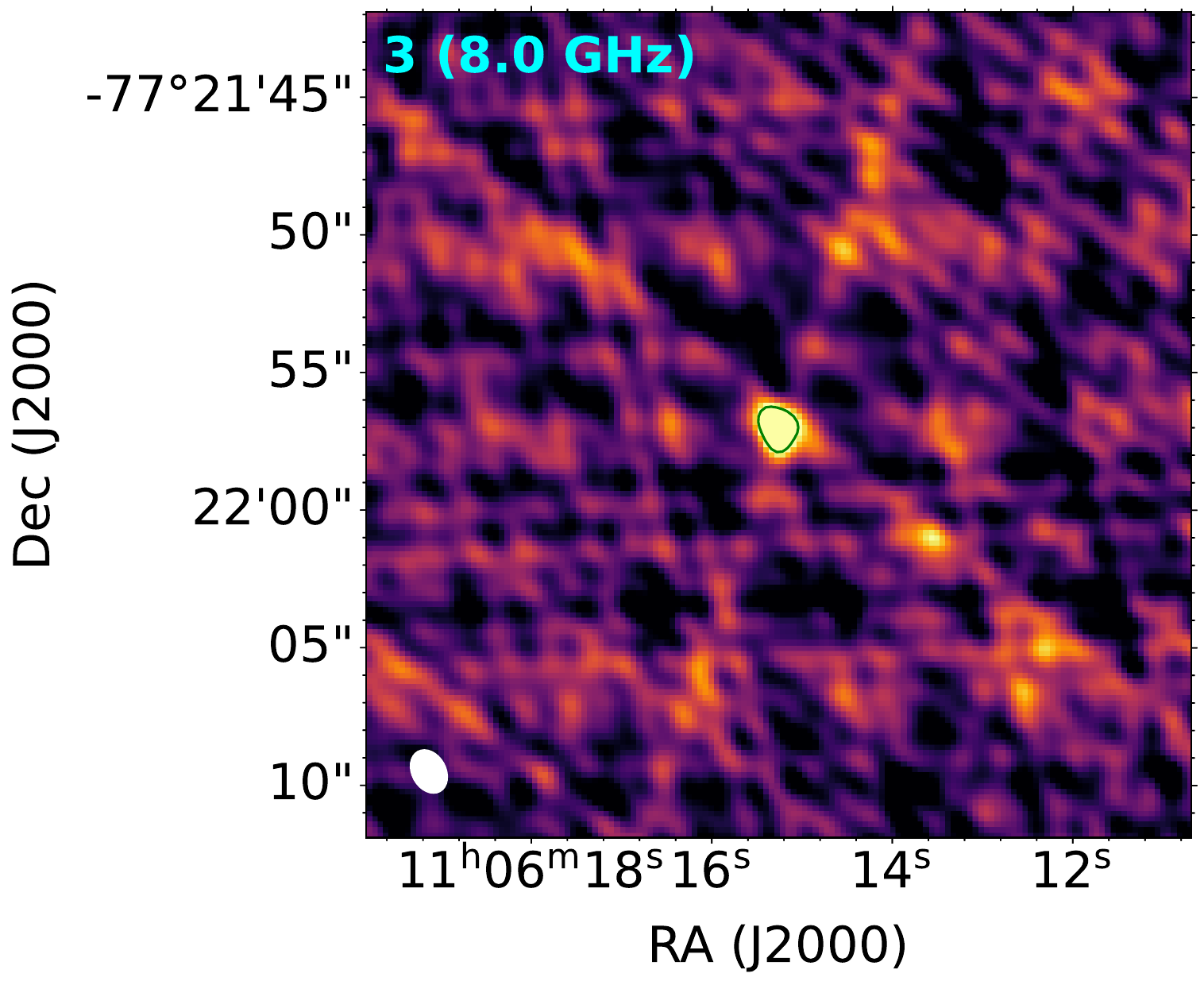}
    \caption{(top) Image of one of the ATCA field in Cha\,IS observed at 5.5 GHz. The small white squares indicate locations of known young stars (from \citealt{Esplin+2017}). Three compact radio sources are clearly seen and are labeled 1, 2, and 3. Zooms on each of these sources (showing the areas marked by cyan squares in the top panel) are shown in the bottom row. The last figure in the bottom row to the right shows source 3 observed at 8 GHz. The contour levels are at 0.2, 0.4, 0.8, 1.6, 3.2, and 6.4~mJy~beam$^{-1}$. The first contour roughly corresponds to $5\sigma$ in all cases. The synthesized beam is shown at the top left corner of each panel.}
    \label{fig:big_image}
\end{figure*}

\input{table2}

\subsection{LBA Observations} \label{sec:lba}
The three most promising sources detected with ATCA (2MASS\,J11061540$-$7721567 and 2MASS\,J11141565$-$7627364 in Cha\,I and 2MASS\,J13005534$-$7708296 in Cha\,II; see Section \ref{sec:lbares}) were observed with the Australian Long Baseline Array (LBA) under project code V329. Four observing sessions were scheduled between October 2016 and September 2018 (Table \ref{tab:LBA_obslog}). All observations were made at a frequency of 8.4 GHz, and used the Australia Telescope National Facility (ATNF) observatories at Parkes (64-m single dish), the Australia Telescope Compact Array (ATCA; five 22-m antennas phased), and Mopra (22-m single dish), as well as the Hobart (26-m single dish) and Ceduna (30-m single dish) antennas operated by the University of Tasmania. In addition, the 70-m (DSS-43) and 34-m antennas (DSS-34 and DSS-36) at Tidbinbilla participated in two epochs. Also, the stations at Hartebeesthoek (26-m single dish) in South Africa, Warkworth (12-m and 30-m antennas) in New Zealand, and Katherine (12-m single dish) and Yarragadee (12-m single dish) in Australia participated in some epochs (see Table \ref{tab:LBA_obslog}). The data were recorded in dual polarization mode (except for DSS-34 and DSS-43 at Tidbinbilla, which recorded only right circular polarization) with 64~MHz of bandwidth in each polarization. The data were correlated at Curtin University using the DiFX software correlator \citep{Deller+2011}. 

The gain calibrators were 1057--797 for the targets in Cha\,I and J1312$-$7724 for the target in Cha\,II. The data reduction was performed using the AIPS software package \citep{AIPS2003} following  standard procedures for phase-referencing observations. First, we applied off-source flags to account for the slewing times of the antennas. Then, we applied corrections for the digital sampling effects of the correlator and for antenna parallactic angle variations. Amplitude calibration was performed by using nominal values of the System Equivalent Flux Density (SEFD) taken from the LBA wiki\footnote{https://www.atnf.csiro.au/vlbi/dokuwiki/doku.php/lbaops/\\lbacalibrationnotes/nominalsefd}. The instrumental single-band delays caused by the antenna electronics were determined from a single scan on a strong calibrator (0637--752, 1343--601 or 1921--293) and then applied to the data. Finally, frequency and time-dependent residual phase errors were obtained by fringe-fitting the phase calibrator data. Once the calibration tables were applied to the data, we produced images of the targets using the CLEAN algorithm (AIPS task IMAGR). We used natural weighting (robust parameter = 5 in IMAGR) in the first epoch and partial uniform (robust = 0) in the last epochs. The rms noise levels of the images and synthesized beam sizes are given in Table \ref{tab:LBA_obslog}. Note that the second epoch (V329 F) was severely affected by bad weather conditions, which resulted in a higher noise.   

\input{table3}

\section{Results} \label{sec:results}

\subsection{ATCA results} \label{sec:atcares}

The 5.5 GHz image of a representative ATCA field in Cha\,IS is shown in Figure \ref{fig:big_image}, where the locations of known young stars are also shown. Three sources are clearly detected in the field as indicated. Two do not coincide with known young stars, while the third one (labeled 3) does. The zooms at the bottom of Figure \ref{fig:big_image} reveal the structure of each source in more detail. Two of them (including the source associated with a young star) are featureless; the third one is clearly elongated in one direction. Figure \ref{fig:herschel} also reveals that the background noise level is not entirely structureless; rather, it shows artifacts caused by the sparse coverage of the ATCA array. The characteristics described here for one specific field in Cha\,IS are generally valid for our ATCA observations.

Table \ref{tab:ATCA_obslog} shows that the noise level in the 5.5 GHz maps is typically between 30 and 40 $\mu$Jy beam$^{-1}$ with a mean value of almost exactly 35 $\mu$Jy beam$^{-1}$. The mean synthesized beam is 3.09$''\times$1.83$''$ at 5.5 GHz so, given the size of the mapped area in each image (a circular area with a radius of 6.6 arcmin), there is a total of $1.8\times10^6$ independent resolution elements in our 28 maps.\footnote{We should emphasize that a resolution element is a synthesized beam (not an individual pixel). Since the maps are over-sampled, with a pixel size at most five times smaller than the major axis of the synthesized beam, adjacent pixels are correlated with one another, so one should not consider individual pixels as independent measurements} By setting a lower limit of $5\sigma$ (0.18 mJy beam$^{-1}$) to consider a feature as a possible detection, we minimize the possibility of contamination by noise peaks, because the probability of such a noise peak above $5\sigma$ is only one in 3.5 million. As a consequence, if the Gaussian noise condition assumed in these calculations held for our observations, we would expect only 0.5 ($= 1.8\times10^5 / 3.5\times10^6$; less than one) noise peaks above $5\sigma$ in the entire surveyed area. Even considering that the noise level is not entirely Gaussian (Figure \ref{fig:big_image}), we expect no contamination above this $5\sigma$ threshold. The average noise level at 8 GHz is slightly higher (45 $\mu$Jy beam$^{-1}$) and the number of independent resolution elements is $2.3\times10^6$, and so the same calculations lead us to expect only 0.65 noise peak above $5\sigma = 0.23$ mJy beam$^{-1}$ in the entire area mapped at 8 GHz.

\input{table4}

Accounting for the above considerations, we systematically searched for peaks above 0.18 mJy beam$^{-1}$ in the 5.5 GHz maps and above 0.23 mJy beam$^{-1}$ in the 8 GHz maps, and found a total of 91 distinct sources.\footnote{Some sources are detected at both frequencies, and some are detected multiple times in overlapping fields (see Figure \ref{fig:herschel}). In such cases they are only counted once.} The great majority of the sources are {\bf not} coincident with known young stars in the region. Some are clearly radio galaxies with resolved lobes, while many are point sources presumably associated with distant quasars. \citet{Fomalont+1991} has shown that the number $N$ of extragalactic sources expected per square arc-minute at 5 GHz above a flux density $S$ is:

\[ \left( \frac{N}{\text{arcmin}^2} \right) = 0.42 \pm 0.05 \left( \frac{S}{30~ \mu\text{Jy}}\right)^{-1.18 \pm 0.19}.\]

\noindent
Given our total surveyed area and $5\sigma$ detection limit, we expect $90\pm30$ extragalactic sources in our observations, fully consistent with the measurements. This shows that almost all sources detected with ATCA are extragalactic with at most a small possible excess from young stars. Indeed, only five of the sources detected above $5\sigma$ coincide with known young stars in the region. Two of them are located in Cha\,IN, two are in Cha\,IS, and one is in Cha\,II (Table \ref{tab:ATCA_YSO}).

The $5\sigma$ cutoff used above is appropriate in a blind search through the entire area surveyed, but it can be significantly relaxed when considering a smaller area. It is well known that, for a normal distribution, there is only a 0.3\% chance of finding a value outside of $3\sigma$. There are 201 young stars located within our ATCA fields and, given the combined astrometric errors of ATCA and typical optical/IR observations, it is sufficient to search the radio maps for counterparts within a circular area of diameter 5 arcsec around each stellar position. This corresponds to a total number of 523 independent resolution elements, so we expect at most one or two $3\sigma$ noise peak feature coincident (within 5 arcsec) with any of the 201 known young stars in the mapped area. The observed images reveal five $3\sigma$ peaks at the position of known young stars (in addition to the five peaks above 5$\sigma$ identified in our blind search); indeed, two such peaks are clearly seen in Figure \ref{fig:big_image} around RA(J2000) = $11^h06^m45^s$. We report these sources in Table \ref{tab:ATCA_YSO} as candidate radio young stars in Chamaeleon. 

\begin{figure*}
    \centering
    \includegraphics[width=12cm]{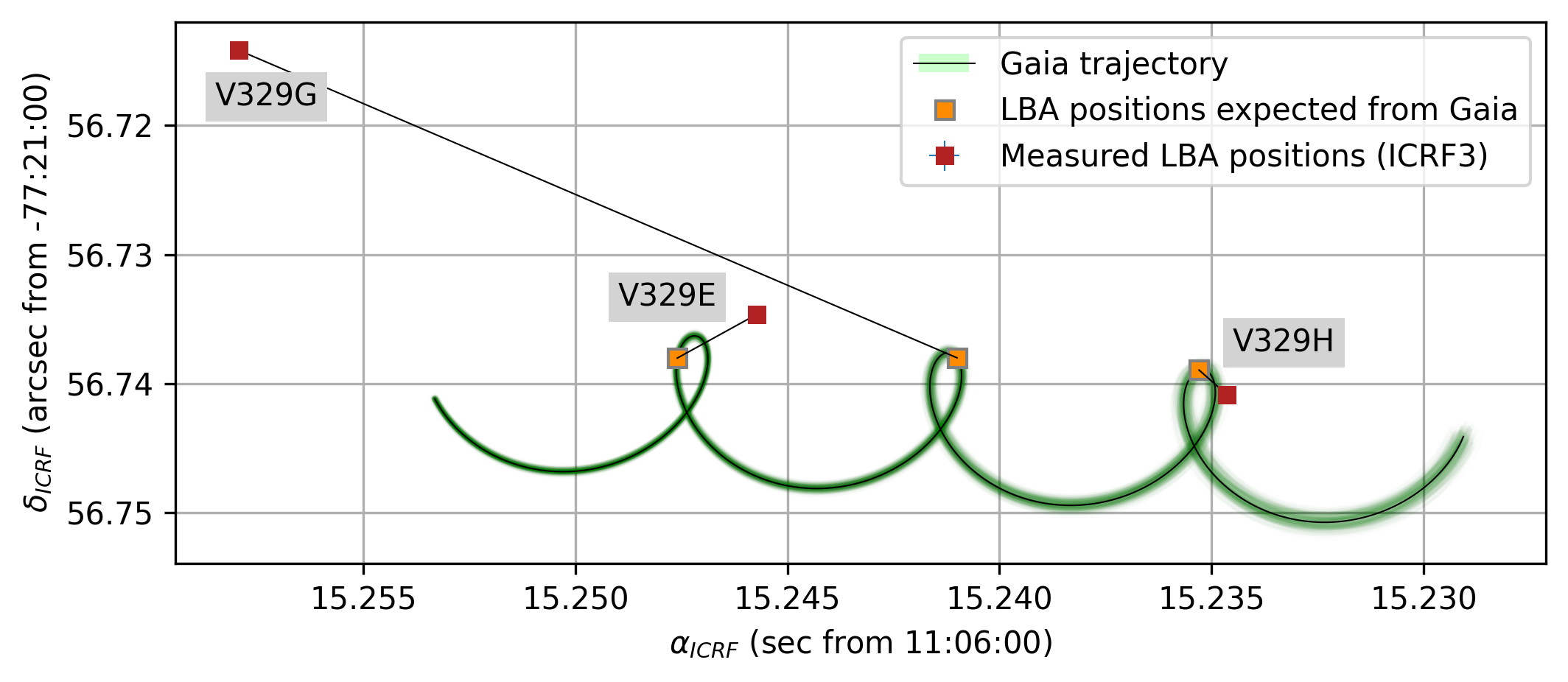}
    \caption{Measured LBA positions for 2MASS\,J11061540$-$7721567 at the three detected epochs and predicted Gaia positions derived from DR3 astrometric elements for that source.}
    \label{fig:astrometry}
\end{figure*}

\subsection{LBA results} \label{sec:lbares}
As described in Section \ref{sec:lba}, the three most promising targets detected with ATCA (2MASS\,J11061540$-$7721567 and 2MASS\,J11141565$-$7627364 in Cha\,IS, and 2MASS\,J13005534$-$7708296 in Cha\,II) were subsequently observed with the LBA. All three sources were observed during the first three epochs (see Table \ref{tab:LBA_obslog}). However, only 2MASS\,J11061540$-$7721567 was detected (in the first two epochs) so we focused on that sole target during the last observation. In total, this led to three LBA detections of 2MASS\,J11061540$-$7721567 as listed in Table \ref{tab:LBA_results} and shown in Figure \ref{fig:astrometry}. The flux density of 2MASS\,J11061540$-$7721567 measured with the LBA, between 0.5 and 0.9 mJy, is moderately variable and comparable to the value found with ATCA ($\sim$ 0.7 mJy). 

Two important points must be mentioned here regarding the LBA positions and their errors. First, in radio interferometric observations (both conventional, such as with ATCA, and long baseline, as with the LBA), the astrometry is referenced to the complex gain calibrator -- 1057--797 in our observations of 2MASS\,J11061540$-$7721567. The cataloged positions of these calibrators have been refined over time using dedicated observations \citep[e.g.,][]{Beasley+2002}. For 1057--797, the cataloged position used during our LBA observations  differs by about 3 mas (mostly along the right ascension direction) from the most recent value in the ICRF3 \citep[][see Table \ref{tab:1057-797}]{Charlot+2020}. This astrometric correction has been included in the measured positions reported in Table \ref{tab:LBA_results} and Figure \ref{fig:astrometry}. We note that 1057--797 was detected by the Gaia satelite and is reported in DR3 with identifier 5199010289211239424 (Table \ref{tab:1057-797}). This enabled us to also verify the alignment between the radio ICRF3 frame and the Gaia optical frame in the region of Chamaeleon. The offset between the radio and Gaia positions of 1057--797 (Table \ref{tab:1057-797}) is $\Delta \alpha = 34~\mu$as, $\Delta \delta = 58~\mu$as, showing that there is excellent agreement between the two reference frames in that portion of the sky. The second point worth mentioning is that the uncertainties on the source positions reported in Table \ref{tab:LBA_results} include a systematic contribution to account for the errors introduced by extrapolating the gains determined on the complex gain calibrator (1057--797) to the target. The angular separation between J11061540$-$7721567 and 1057--797 is 2.7$^{\circ}$; following \citet{Reid+2017}, we assume a systematic error of 0.1 mas per degree of separation and therefore considered a systematic error of 0.27 mas along each direction.

\section{Discussion} \label{sec:discus}

\begin{figure}
    \makebox[\textwidth][l]{
    \hspace{-1.0cm}
    \includegraphics[width=10cm]{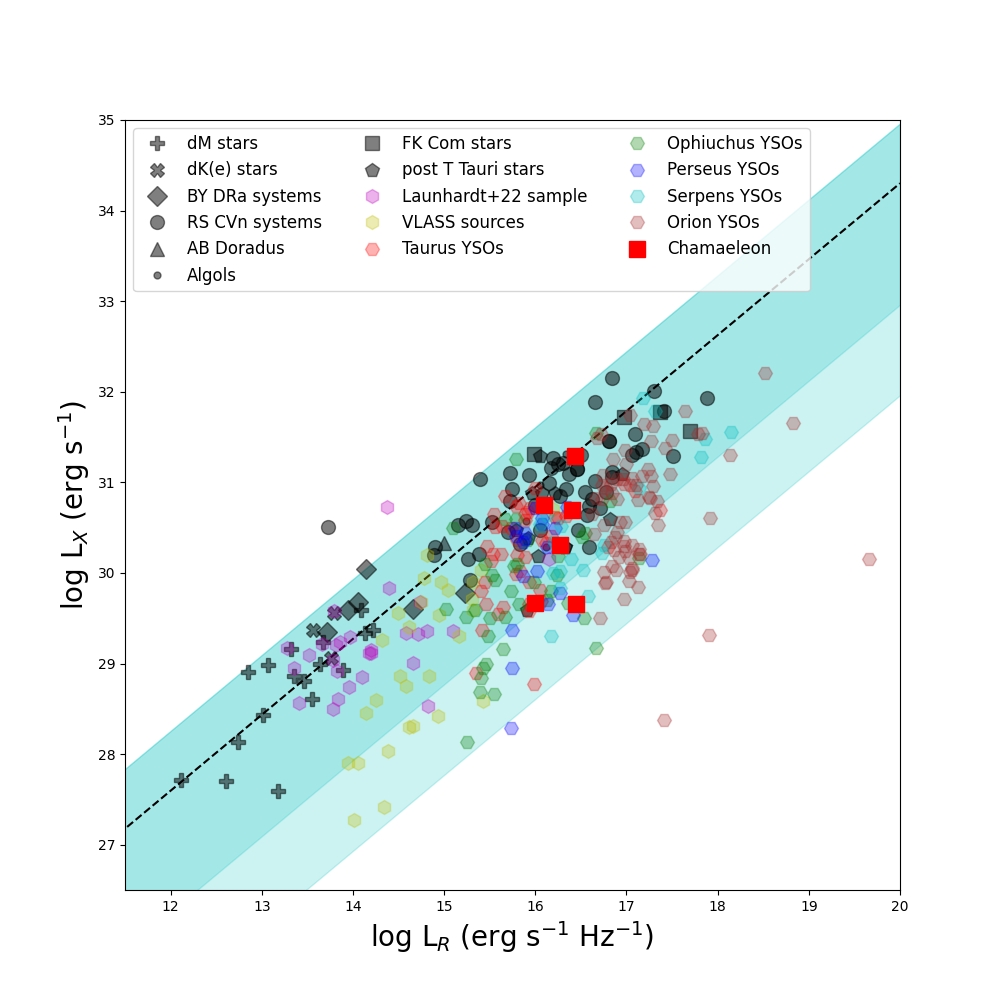}
    }
    \caption{Radio emission as a function of X-ray luminosity for a sample of stellar sources. The grey symbols correspond to the original data set used by \citet{Gudel1993}. The YSO samples are from \citet{Dzib+2013}, \citet{Kounkel+2014}, \citet{OrtizLeon+2015}, \citet{Dzib+2015}, and \citet{Pech+2016}. Additionally, we included the data from \citet{Launhardt+2022} and \citet{Yiu+2024}. The six stars reported here (Table \ref{tab:ATCA_YSO}) as definite or candidates radio stars that have known X-ray counterparts, from \citealt{Stelzer+2004} and \citealt{Feigelson+2004}, are shown as red squares. The dashed line ($\log L_X = 0.84 \log L_R+17.54$) shows a fit to dM stars. The darker cyan band shows a range from $-1.34$ to $+0.66$ dex from the dM star fit, while the lighter cyan band extends down to $-2.34$ dex.}
    \label{fig:xray}
\end{figure}

Our ATCA observations enabled us to firmly detect radio emission toward five young stars in Chamaeleon and obtain five additional tentative detections. Since the surveyed region includes a total of 201 young stars, the detection rate is between 2.5 and 5\%.  This is somewhat lower than in other star-forming regions (e.g., Ophiuchus; \citealt{Dzib+2013}) but comparable to others (such as Perseus; \citealt{Pech+2016}). When they could be measured, the radio spectral indices found here in Chamaeleon are mostly comparable with those reported in other regions \citep{Dzib+2013}, with typical values that are slightly positive. One source, however, exhibits a very negative spectral index clearly indicative of a non-thermal emission process. As in other regions, most detected sources are associated with (often eruptive) young stars of spectral types M and K (Table \ref{tab:ATCA_YSO}), but two interesting cases should be mentioned. One is our definite detection of a Herbig Ae/Be star (spectral type B9) and the other one is the tentative detection of radio emission associated with one of the few protostellar sources in Chamaeleon (Ced\,110~IRS4). The former adds to a limited number of known intermediate-mass young stars with radio emission \citep[e.g.,][]{Dzib+2010,Jazmin+2024}. The latter is certainly not unexpected since embedded protostars tend to power partially ionized jets that can be powerful radio sources \citep{Anglada+2018}. The existence of a collimated jet in Ced\,110~IRS4 has indeed recently been confirmed by JWST observations \citep{Narang2025}. In that specific case, then, the radio emission mechanism is likely thermal brehmsstrahlung. The other detected objects are associated with fairly evolved low-mass young stars (Class\,III and/or WTTS) so the emission process is likely non-thermal. This is consistent with Figure \ref{fig:xray} where we compare the X-ray and radio properties of the sources detected in Chamaeleon with those of a sample of other active stars and show that the Chamaeleon sources follow the empirical G\"udel-Benz relationship \citep{Gudel1993}.

Three of our definite detections have previously been reported in the literature. 2MASS\,J11061540$-$7721567 (also known as Ced\,110~IRS2) and 2MASS\,J11075588$-$7727257 (CHX\,10a) were detected by both \citet{Lehtinen+2003} and \citet{Brown+1996}, while 2MASS\,J11095003$-$7636476 (HD\,97300) was reported by \citet{Brown+1996}. In addition, two of our candidate detections, Ced\,110~IRS4 and 2MASS\,J11091769$-$7627578 (CHXR\,37) were previously reported by \citet{Lehtinen+2003} and \citet{Brown+1996}, respectively. The remaining two definite and three candidate detections are reported here for the first time. We note that \citet{Brown+1996} report on two additional radio detections: 2MASS\,J11100010$-$7634578 (WW\,Cha) and 2MASS\,J11054300$-$7726517 (CHXR\,15), so there are now twelve stellar sources in Chamaeleon with confirmed or candidate radio detections. On the basis of the stellar and radio properties of the detected targets, \citet{Lehtinen+2003} argue in favor of a thermal emission mechanism for Ced\,110~IRS2 and Ced\,110~IRS4, and a non-thermal process in CHX10a. We reached the same conclusions in the previous paragraph for Ced\,110~IRS4 and CHX10a, but argued in favor of a non-thermal mechanism in Ced\,110~IRS2 given that this object is a fairly evolved young star. We will see momentarily that the LBA observations confirmed that conclusion. \citet{Brown+1996} briefly discuss their detections in the context of active stellar coronae, also implying a non-thermal origin.

\input{table5}

Only one (2MASS\,J11061540$-$7721567, Ced\,110~IRS2) of the three brightest sources detected with ATCA was detected on long baselines with the LBA, with the positions and fluxes reported in Table \ref{tab:LBA_results}. The LBA detection unequivocally established the non-thermal nature of the emission from that specific target because very long baseline arrays are only sensitive to emission with a brightness temperature above about $10^6$ K \citep{Moran+2017}. The source positions change appreciably in the different epochs, also confirming that 2MASS\,J11061540$-$7721567 is a Galactic source (rather than a background extragalactic source coincidentally located in the exact direction of a young star in Chamaeleon\footnote{This was already an exceedingly low probability, but is fully ruled out with the LBA results.}).

2MASS\,J11061540$-$7721567 was detected with Gaia under DR3 id 5201203647406601600. Figure \ref{fig:astrometry} shows the astrometric trace of 2MASS\,J11061540$-$7721567 expected from its Gaia DR3 astrometric elements together with the measured LBA positions. For the first and last epochs (E and H), the agreement between the Gaia and LBA positions is good, but there is a large offset, of order 60 mas, for the second epoch (G). Such a large offset cannot be attributed to astrometric errors and most likely indicates that Gaia and the LBA detect the same object at epochs E and H, but not at epoch G. This would imply that J11061540$-$7721567 is a binary system, with the radio emission tracing a different star at different epochs. The binarity of this target could actually be suspected from the Gaia data alone because the Renormalised Unit Weight Error (RUWE) parameter that Gaia uses to quantify the quality of a single-star astrometric fit is large (9.02) indicating significant post-fit residuals. The recommendation from the Gaia consortium is to use RUWE $<$ 1.4 as a limiting value for high quality single-star fits. That the two stars in a given binary system can turn on and off as radio sources could, at first, sound unlikely, but such a behavior has been documented in several other young stellar systems \citep[e.g.,][]{OrtizLeon+2017,Jazmin+2024}.

With only one LBA detection of the second star in 2MASS\,J11061540$-$7721567, it is evidently impossible to model the orbit (at least half a dozen additional data points would be needed to attempt a full astrometric fit). We note, however, that the separation between the LBA and the expected Gaia positions for epoch G, about 60 mas, corresponds to about 12 AU at the distance of Chamaeleon. A 1 $M_\odot$ binary system with such a semi-major axis would have an orbital period of about 40 years, so monitoring the system with the LBA over the next few decades ought to enable a good determination of the orbital elements and the mass of the individual stars in 2MASS\,J11061540$-$7721567.

\section{Conclusions} \label{sec:conclusions}
Using large-scale high-resolution observations of the Chamaeleon star-forming region with ATCA, we have detected radio emission from five young stars and obtained tentative detections from five more. Except for one protostellar source (Ced\,110~IRS4), the detected young stars are fairly evolved T Tauri stars and the radio emission mechanism is likely non-thermal. Given the known radio variability of young stars, it would be very useful to repeat ATCA observations similar to those reported here to monitor the detected targets and potentially detect more radio-bright stars that may have been in a low state during our observing run.

One of the brightest stars detected with ATCA (2MASS\,J11061540$-$7721567, Ced\,110~IRS2) was also detected on very long baselines with the LBA. Comparison between the Gaia DR3 and LBA astrometry reveals a large offset (about 60 mas) for one epoch, strongly suggesting that 2MASS\,J11061540$-$7721567 is in fact a tight binary system with an orbital period of order 40 years. This conclusion could already be expected from Gaia results alone given that the Gaia astrometry fit with a single-star model produced large residuals. Additional LBA observations will be required to constrain the orbital motions of the system. Systematic LBA observations of all confirmed and candidate radio detections of young stars in Chameleon would also be useful to constrain the overall astrometry of the region.

\section*{Acknowledgements}

The Australia Telescope Compact Array and the Long Baseline Array are part of the Australia Telescope National Facility (https://ror.org/05qajvd42) which is funded by the Australian Government for operation as a National Facility managed by CSIRO. This work was supported by resources provided by the Pawsey Supercomputing Research Centre with funding from the Australian Government and the Government of Western Australia. L.L. acknowledges the support of DGAPA PAPIIT grants IN112416, IN108324 and IN112820 as well as CONAHCyT-CF grant 263356. S.A.D. acknowledges the M2FINDERS project from the European Research Council (ERC) under the European Union's Horizon 2020 research and innovation programme (grant No 101018682). We thank Jacopo Fritz for his assistance with the Herschel data, Chikaedu Ogbodo for his help with collecting data at the LBA, and the referee, Fabrizio Massi, for a thoughtful report that helped improve the quality and clarity of the paper.

\section*{Data Availability}

All data included in this paper are available through the ATCA and LBA data archive systems.

\bibliographystyle{mnras}
\bibliography{ms} % if your bibtex file is called example.bib

\label{lastpage}
\end{document}

%% file: table1.tex
\begin{table*}
\centering
\caption{ATCA observation logs and image properties}
\label{tab:ATCA_obslog}
\begin{tabular}{cccllccc}
\hline
Region & RA(J2000) & Dec(J2000) & Beam @ 5.5 GHz & Beam @ 8.0 GHz & $\sigma_{5.5 \text{~GHz}}$ & $\sigma_{8.0 \text{~GHz}}$ \\
       & {$^h : ^m : ^s$} & {$^\circ : ' : ''$} & \multicolumn{2}{c}{$\theta_{\text{max}} ('') \times \theta_{\text{min}} ('')$; PA($^\circ$)}  &  \multicolumn{2}{c}{$\mu$Jy beam$^{-1}$} \\
\hline
 1N & 11:04:37 & $-$76:26:42 & 3.52$\times$1.78; 29.8 & 2.23$\times$1.12; 29.2 & 36.4 & 52.7 \\ %F12
 1N & 11:06:54 & $-$76:26:00 & 3.53$\times$1.78; 32.1 & 2.23$\times$1.12; 31.9 & 35.5 & 46.6 \\ %F15
 1N & 11:06:54 & $-$76:33:50 & 3.52$\times$1.78; 32.8 & 2.23$\times$1.13; 33.1 & 34.9 & 45.0 \\ %F16
 1N & 11:09:13 & $-$76:39:54 & 3.52$\times$1.78; 29.8 & 2.23$\times$1.12; 29.2 & 40.3 & 54.3 \\ %F13
 1N & 11:09:40 & $-$76:24:26 & 3.52$\times$1.78; 33.1 & 2.23$\times$1.22; 38.0 & 33.6 & 46.2 \\ %F17
 1N & 11:09:42 & $-$76:31:53 & 2.59$\times$1.89; 38.7 & 1.64$\times$1.22; 40.0 & 33.3 & 44.2 \\ %F10
 1N & 11:11:45 & $-$76:44:49 & 3.53$\times$1.79; 33.7 & 2.22$\times$1.13; 34.4 & 34.3 & 45.8 \\ %F18
 1N & 11:12:25 & $-$76:35:58 & 2.59$\times$1.89; 37.4 & 1.65$\times$1.22; 38.0 & 37.1 & 42.4 \\ %F09
 1N & 11:13:54 & $-$76:27:32 & 2.58$\times$1.89; 38.9 & 1.65$\times$1.22; 40.0 & 38.8 & 44.6 \\ %F11
\\[-6.5pt]
 1S & 11:02:01 & $-$77:31:55 & 2.58$\times$1.89; 38.2 & 1.64$\times$1.22; 35.7 & 39.4 & 46.4 \\ %F19
 1S & 11:03:29 & $-$77:23:29 & 2.59$\times$1.89; 32.0 & 1.65$\times$1.21; 32.0 & 34.5 & 42.5 \\ %F05
 1S & 11:06:19 & $-$77:24:37 & 2.59$\times$1.88; 29.3 & 1.65$\times$1.22; 28.9 & 32.9 & 40.1 \\ %F03
 1S & 11:07:22 & $-$77:36:59 & 2.59$\times$1.88; 29.9 & 1.65$\times$1.22; 29.7 & 31.6 & 41.9 \\ %F04
 1S & 11:07:36 & $-$77:43:44 & 2.59$\times$1.88; 27.7 & 1.65$\times$1.22; 27.8 & 35.1 & 38.7 \\ %F02
 1S & 11:07:54 & $-$77:17:35 & 2.58$\times$1.88; 34.8 & 1.65$\times$1.22; 35.1 & 32.9 & 41.7 \\ %F08
 1S & 11:08:37 & $-$77:29:51 & 2.59$\times$1.89; 32.5 & 1.65$\times$1.21; 32.6 & 30.8 & 40.7 \\ %F06
 1S & 11:09:59 & $-$77:36:58 & 2.59$\times$1.89; 33.1 & 1.65$\times$1.22; 33.6 & 31.5 & 39.7 \\ %F07
 1S & 11:10:53 & $-$77:26:57 & 2.58$\times$1.88; 35.1 & 1.65$\times$1.22; 35.4 & 33.4 & 45.2 \\ %F14
 1S & 11:11:10 & $-$77:20:07 & 2.59$\times$1.88; 25.7 & 1.65$\times$1.22; 26.1 & 30.0 & 39.5 \\ %F01
\\[-6.5pt]
 2  & 12:57:00 & $-$76:44:00 & 3.53$\times$1.78; 14.4 & 2.24$\times$1.12; 14.4 & 33.9 & 47.0 \\ %F05
 2  & 13:00:20 & $-$77:11:20 & 3.53$\times$1.76; 13.0 & 2.24$\times$1.12; 12.8 & 37.5 & 46.5 \\ %F04
 2  & 13:03:21 & $-$77:53:17 & 3.51$\times$1.77; 16.7 & 2.23$\times$1.12; 17.2 & 37.8 & 47.5 \\ %F08
 2  & 13:05:13 & $-$77:39:37 & 3.53$\times$1.78; 10.0 & 2.24$\times$1.12; 10.8 & 36.9 & 47.3 \\ %F03
 2  & 13:06:15 & $-$77:32:15 & 3.52$\times$1.78; 8.7  & 2.24$\times$1.12; 9.2  & 35.8 & 47.4 \\ %F02
 2  & 13:06:59 & $-$77:23:31 & 3.52$\times$1.77; 15.3 & 2.24$\times$1.12; 14.8 & 39.1 & 48.4 \\ %F07
 2  & 13:07:58 & $-$77:39:54 & 3.52$\times$1.78; 7.7  & 2.24$\times$1.12; 7.5  & 39.4 & 46.7 \\ %F01
 2  & 13:09:01 & $-$77:56:12 & 3.53$\times$1.76; 13.6 & 2.25$\times$1.12; 13.5 & 36.1 & 53.2 \\ %F06 
 2  & 13:09:33 & $-$77:10:23 & 3.52$\times$1.77; 16.3 & 2.24$\times$1.12; 16.8 & 39.3 & 47.4 \\ %F09
\hline
\end{tabular}
\end{table*}

%% file: table2.tex
\begin{table*}
\centering
\caption{LBA observation logs and image properties}
\label{tab:LBA_obslog}
\begin{tabular}{cccccc}
\hline
Obs code & Date         & LBA Stations{$\dagger$}  & 2MASS Target  & Beam & $\sigma$ \\
         & yyyy-mm-dd   &                                        &               & $\theta_{\text{max}} \text{~(mas)} \times \theta_{\text{min}} \text{~(mas)}$; PA ($^\circ$)     & ($\mu$Jy bm$^{-1}$)\\
\hline
V329\,E & 2016-10-18               & {AT, CD, HO, MP,}           & J11061540$-$7721567 & 10.06$\times$7.98; $-$89.1 & 16.9    \\
        &                          & {PA, TB, TI}                & J11141565$-$7627364 & 10.01$\times$7.89; $+$88.3 & 16.0    \\
        &                          &                             & J13005534$-$7708296 & 10.68$\times$7.56; $+$75.5 & 20.1    \\        
\\[-0.3cm]
V329\,F & 2017-01-19               & {AT, CD, HH, HO,}           & J11061540$-$7721567 & 2.75$\times$1.60; $+$26.8  & 164.5 \\
        &                          & {MP, VW}                    & J11141565$-$7627364 & 2.42$\times$0.43; $+$26.8  & 156.5 \\
        &                          &                             & J13005534$-$7708296 & 3.24$\times$1.31; $+$7.4   & 357.3 \\        
\\[-0.3cm]
V329\,G & 2017-08-10               & {AT, CD, HO, MP,}           & J11061540$-$7721567 & 2.22$\times$2.18; $+$69.4  & 67.1    \\
        &                          & {PA, WA}                    & J11141565$-$7627364 & 2.21$\times$2.11; $+$18.4  & 75.6    \\
        &                          &                             & J13005534$-$7708296 & 2.10$\times$1.94; $-$57.7  & 112.5   \\  
\\[-0.3cm]
V329\,H & 2018-09-08               & {AT, HO, KE, MP, }          & J11061540$-$7721567 & 2.81$\times$2.40; $+$31.8  & 30.8    \\%
        &                          & {PA, TD, TI, WA, YG}\\
\hline
\end{tabular}\\
$\dagger$ Station codes: AT = ATCA; CD = Ceduna; HO = Hobart; MP = Mopra; PA = Parkes; TB = Tidbinbilla DSS-34; TI = Tidbinbilla DSS-43; TD = Tidbinbilla DSS-36; HH = Hartebeesthoek; WW = Warkworth 12m; WA = Warkworth 30m; KE = Katherine; YG = Yarragadee.
\end{table*}

%% file: table3.tex
\begin{table*}
\centering
\caption{Radio young stars in Chamaeleon detected with ATCA}
\label{tab:ATCA_YSO}
\begin{tabular}{llcccccc}
\hline
RA(J2000) & Dec(J2000) & $\nu$ & $S_\nu$ & $\alpha$ & Region & 2MASS Name  & Spectral \\
          &            & (GHz) & (mJy)   &          &        &             & Type{$\dagger$}     \\
\hline
\multicolumn{2}{l}{\bf Definite detections:}\\
11:06:15.250(7)     & $-$77:21:56.95(4)    & $5.5$           & $0.66\pm0.04$     & $+0.23\pm0.27$     & Cha\,IS          & J11061540$-$7721567  & G5              \\%cha1s-f03
                    &                      & $8.0$           & $0.72\pm0.06$     &                    &  \\
\\[-0.3cm]
11:07:56.107(61)    & $-$77:27:26.13(10)   & $5.5$           & $0.61\pm0.08$     & --                 & Cha\,IS          & J11075588$-$7727257  & K6              \\%cha1s-f06
\\[-0.3cm]
11:09:49.899(24)    & $-$76:36:47.45(14)   & $5.5$           & $0.64\pm0.08$     & --                 & Cha\,IN          & J11095003$-$7636476  & B9               \\%cha1n-f13
\\[-0.3cm]
11:14:15.567(7)     & $-$76:27:36.43(3)    & $5.5$           & $0.73\pm0.03$     &  $-1.74\pm0.44$    & Cha\,IN          & J11141565$-$7627364  & M3.75              \\%cha1n-f11
                    &                      & $8.0$           & $0.38\pm0.06$     &                    & \\
\\[-0.3cm]
13:00:55.225(6)     & $-$77:08:29.77(6)    & $5.5$           & $2.09\pm0.06$     & $+0.05\pm0.35$     & Cha\,II          & J13005534$-$7708296   & M2.5              \\%cha2-f04
                    &                      & $8.0$           & $2.13\pm0.27$     &                    &  \\
\multicolumn{2}{l}{\bf Candidate detections:}\\
11:06:43.426(63)    & $-$77:26:35.18(27)   & $5.5$           & $0.24\pm0.08$     & $+0.41\pm1.17$     & Cha\,IS          & J11064346$-$7726343   & M3               \\%cha1s-f03
                    &                      & $8.0$           & $0.28\pm0.08$\\
\\[-0.3cm]
11:06:46.427(22)    & $-$77:22:33.39(11)   & $5.5$           & $0.26\pm0.04$     & --                 & Cha\,IS          & Ced\,110~IRS4         &                   \\%cha1s-f03
\\[-0.3cm]
11:08:01.482(43)    & $-$77:42:28.79(13)   & $5.5$           & $0.45\pm0.06$     & --                 & Cha\,IS          & J11080148$-$7742288   & K8                \\%cha1s-f02
\\[-0.3cm]
11:09:11.548(95)    & $-$77:29:13.05(14)   & $5.5$           & $0.30\pm0.08$     & --                 & Cha\,IS          & J11091172$-$7729124   & M2               \\%cha1s-f06
                    &                      &                 &                   &                    &                  & J11091171$-$7729120   & M3               \\
\\[-0.3cm]
11:09:17.471(22)    & $-$76:27:58.23(8)    & $5.5$           & $0.35\pm0.04$     & --                 & Cha\,IN          & J11091769$-$7627578   & K7             \\%cha1n-f17
\hline
\end{tabular}\\
$\dagger$ The spectral types are taken from \citet{Alcala+2008} for Cha\,I and \citet{Spezzi+2008} for Cha\,II.
\end{table*}

%% file: table4.tex
\begin{table*}
\centering
\caption{Positions and flux densities for the three LBA detections of 2MASS\,J11061540$-$7721567}
\label{tab:LBA_results}
\begin{tabular}{cccccc}
\hline
Obs code & JD data & RA(J2000)  & Dec(J2000)   & $S_\nu$  \\                                                                    
         &                      & $^h : ^m : ^s \pm ^s$  &  $^\circ : ' : '' \pm ''$        & (mJy)\\
\hline
V329\,E & 2457680.38 & 11:06:15.24573 $\pm$ 0.00010 & $-$77:21:56.73465 $\pm$ 0.00038 & 0.48$\pm$0.04\\
V329\,G & 2457976.45 & 11:06:15.25794 $\pm$ 0.00002 & $-$77:21:56.71419 $\pm$ 0.00028 & 0.86$\pm$0.06\\
V329\,H & 2458370.48 & 11:06:15.23464 $\pm$ 0.00005 & $-$77:21:56.74084 $\pm$ 0.00032 & 0.52$\pm$0.07\\
% Before quasar correction
%V329\,E & 2457680.38 & 11:06:15.24627 $\pm$ 0.00014 & $-$77:21:56.73484 $\pm$ 0.00040 & 0.48$\pm$0.04\\
%V329\,G & 2457976.45 & 11:06:15.25848 $\pm$ 0.00011 & $-$77:21:56.71438 $\pm$ 0.00028 & 0.86$\pm$0.06\\
%V329\,H & 2458370.48 & 11:06:15.23518 $\pm$ 0.00011 & $-$77:21:56.74103 $\pm$ 0.00030 & 0.52$\pm$0.07\\
\hline
\end{tabular}
\end{table*}

%% file: table5.tex
\begin{table}
\centering
\caption{Positions of the gain calibrator 1057-797 in various catalogs}
\label{tab:1057-797}
\begin{tabular}{lccc}
\hline
Catalog & RA(J2000)      & Dec(J2000) \\
        & $^h : ^m : ^s$ & $^\circ : ' : ''$\\
\hline
ATCA/LBA & 10:58:43.31100000 & --80:03:54.1600000 \\
ICRF3    & 10:58:43.30979475 & --80:03:54.1598058 \\
Gaia DR3 & 10:58:43.30980797 & --80:03:54.1598637 \\
\hline
\end{tabular}
\end{table}

%% file: ms.bbl
\begin{thebibliography}{}
\makeatletter
\relax
\def\mn@urlcharsother{\let\do\@makeother \do\$\do\&\do\#\do\^\do\_\do\%\do\~}
\def\mn@doi{\begingroup\mn@urlcharsother \@ifnextchar [ {\mn@doi@} {\mn@doi@[]}}
\def\mn@doi@[#1]#2{\def\@tempa{#1}\ifx\@tempa\@empty \href {http://dx.doi.org/#2} {doi:#2}\else \href {http://dx.doi.org/#2} {#1}\fi \endgroup}
\def\mn@eprint#1#2{\mn@eprint@#1:#2::\@nil}
\def\mn@eprint@arXiv#1{\href {http://arxiv.org/abs/#1} {{\tt arXiv:#1}}}
\def\mn@eprint@dblp#1{\href {http://dblp.uni-trier.de/rec/bibtex/#1.xml} {dblp:#1}}
\def\mn@eprint@#1:#2:#3:#4\@nil{\def\@tempa {#1}\def\@tempb {#2}\def\@tempc {#3}\ifx \@tempc \@empty \let \@tempc \@tempb \let \@tempb \@tempa \fi \ifx \@tempb \@empty \def\@tempb {arXiv}\fi \@ifundefined {mn@eprint@\@tempb}{\@tempb:\@tempc}{\expandafter \expandafter \csname mn@eprint@\@tempb\endcsname \expandafter{\@tempc}}}

\bibitem[\protect\citeauthoryear{{Alcala}, {Krautter}, {Schmitt}, {Covino}, {Wichmann}  \& {Mundt}}{{Alcala} et~al.}{1995}]{Alcala+1995}
{Alcala} J.~M.,  {Krautter} J.,  {Schmitt} J.~H.~M.~M.,  {Covino} E.,  {Wichmann} R.,   {Mundt} R.,  1995, \aaps, \href {https://ui.adsabs.harvard.edu/abs/1995A&AS..114..109A} {114, 109}

\bibitem[\protect\citeauthoryear{{Alcal{\'a}}, {Covino}, {Sterzik}, {Schmitt}, {Krautter}  \& {Neuh{\"a}user}}{{Alcal{\'a}} et~al.}{2000}]{Alcala+2000}
{Alcal{\'a}} J.~M.,  {Covino} E.,  {Sterzik} M.~F.,  {Schmitt} J.~H.~M.~M.,  {Krautter} J.,   {Neuh{\"a}user} R.,  2000, \aap, \href {https://ui.adsabs.harvard.edu/abs/2000A&A...355..629A} {355, 629}

\bibitem[\protect\citeauthoryear{{Alcal{\'a}} et~al.,}{{Alcal{\'a}} et~al.}{2008}]{Alcala+2008}
{Alcal{\'a}} J.~M.,  et~al., 2008, \mn@doi [\apj] {10.1086/527315}, \href {https://ui.adsabs.harvard.edu/abs/2008ApJ...676..427A} {676, 427}

\bibitem[\protect\citeauthoryear{{Andr{\'e}} et~al.,}{{Andr{\'e}} et~al.}{2010}]{Andre+2010}
{Andr{\'e}} P.,  et~al., 2010, \mn@doi [\aap] {10.1051/0004-6361/201014666}, \href {https://ui.adsabs.harvard.edu/abs/2010A&A...518L.102A} {518, L102}

\bibitem[\protect\citeauthoryear{{Anglada}, {Rodr{\'\i}guez}  \& {Carrasco-Gonz{\'a}lez}}{{Anglada} et~al.}{2018}]{Anglada+2018}
{Anglada} G.,  {Rodr{\'\i}guez} L.~F.,   {Carrasco-Gonz{\'a}lez} C.,  2018, \mn@doi [\aapr] {10.1007/s00159-018-0107-z}, \href {https://ui.adsabs.harvard.edu/abs/2018A&ARv..26....3A} {26, 3}

\bibitem[\protect\citeauthoryear{{Beasley}, {Gordon}, {Peck}, {Petrov}, {MacMillan}, {Fomalont}  \& {Ma}}{{Beasley} et~al.}{2002}]{Beasley+2002}
{Beasley} A.~J.,  {Gordon} D.,  {Peck} A.~B.,  {Petrov} L.,  {MacMillan} D.~S.,  {Fomalont} E.~B.,   {Ma} C.,  2002, \mn@doi [\apjs] {10.1086/339806}, \href {https://ui.adsabs.harvard.edu/abs/2002ApJS..141...13B} {141, 13}

\bibitem[\protect\citeauthoryear{{Belloche}, {Parise}, {Schuller}, {Andr{\'e}}, {Bontemps}  \& {Menten}}{{Belloche} et~al.}{2011}]{Belloche+2011}
{Belloche} A.,  {Parise} B.,  {Schuller} F.,  {Andr{\'e}} P.,  {Bontemps} S.,   {Menten} K.~M.,  2011, \mn@doi [\aap] {10.1051/0004-6361/201117276}, \href {https://ui.adsabs.harvard.edu/abs/2011A&A...535A...2B} {535, A2}

\bibitem[\protect\citeauthoryear{{Bertout}, {Robichon}  \& {Arenou}}{{Bertout} et~al.}{1999}]{Bertout+1999}
{Bertout} C.,  {Robichon} N.,   {Arenou} F.,  1999, \mn@doi [\aap] {10.48550/arXiv.astro-ph/9909438}, \href {https://ui.adsabs.harvard.edu/abs/1999A&A...352..574B} {352, 574}

\bibitem[\protect\citeauthoryear{{Briggs}}{{Briggs}}{1995}]{Briggs1995}
{Briggs} D.~S.,  1995, PhD thesis, New Mexico Institute of Mining and Technology

\bibitem[\protect\citeauthoryear{{Brown}, {Walter}, {Ambruster}, {Stewart}  \& {Jeffries}}{{Brown} et~al.}{1996}]{Brown+1996}
{Brown} A.,  {Walter} F.~M.,  {Ambruster} C.,  {Stewart} R.~T.,   {Jeffries} R.,  1996, in {Taylor} A.~R.,  {Paredes} J.~M.,  eds,  Astronomical Society of the Pacific Conference Series Vol. 93, Radio Emission from the Stars and the Sun. p.~294

\bibitem[\protect\citeauthoryear{{Brown}, {Day}  \& {Walter}}{{Brown} et~al.}{2004}]{Brown+2004}
{Brown} A.,  {Day} F.,   {Walter} F.~M.,  2004, in American Astronomical Society Meeting Abstracts \#204. p. 62.11

\bibitem[\protect\citeauthoryear{{CASA Team} et~al.,}{{CASA Team} et~al.}{2022}]{CASA2022}
{CASA Team} et~al., 2022, \mn@doi [\pasp] {10.1088/1538-3873/ac9642}, \href {https://ui.adsabs.harvard.edu/abs/2022PASP..134k4501C} {134, 114501}

\bibitem[\protect\citeauthoryear{{Charlot} et~al.,}{{Charlot} et~al.}{2020}]{Charlot+2020}
{Charlot} P.,  et~al., 2020, \mn@doi [\aap] {10.1051/0004-6361/202038368}, \href {https://ui.adsabs.harvard.edu/abs/2020A&A...644A.159C} {644, A159}

\bibitem[\protect\citeauthoryear{{Deller} et~al.,}{{Deller} et~al.}{2011}]{Deller+2011}
{Deller} A.~T.,  et~al., 2011, \mn@doi [\pasp] {10.1086/658907}, \href {https://ui.adsabs.harvard.edu/abs/2011PASP..123..275D} {123, 275}

\bibitem[\protect\citeauthoryear{{Dobashi}, {Uehara}, {Kandori}, {Sakurai}, {Kaiden}, {Umemoto}  \& {Sato}}{{Dobashi} et~al.}{2005}]{Dobashi+2005}
{Dobashi} K.,  {Uehara} H.,  {Kandori} R.,  {Sakurai} T.,  {Kaiden} M.,  {Umemoto} T.,   {Sato} F.,  2005, \mn@doi [\pasj] {10.1093/pasj/57.sp1.S1}, \href {https://ui.adsabs.harvard.edu/abs/2005PASJ...57S...1D} {57, S1}

\bibitem[\protect\citeauthoryear{{Dzib}, {Loinard}, {Mioduszewski}, {Boden}, {Rodr{\'\i}guez}  \& {Torres}}{{Dzib} et~al.}{2010}]{Dzib+2010}
{Dzib} S.,  {Loinard} L.,  {Mioduszewski} A.~J.,  {Boden} A.~F.,  {Rodr{\'\i}guez} L.~F.,   {Torres} R.~M.,  2010, \mn@doi [\apj] {10.1088/0004-637X/718/2/610}, \href {https://ui.adsabs.harvard.edu/abs/2010ApJ...718..610D} {718, 610}

\bibitem[\protect\citeauthoryear{{Dzib} et~al.,}{{Dzib} et~al.}{2013}]{Dzib+2013}
{Dzib} S.~A.,  et~al., 2013, \mn@doi [\apj] {10.1088/0004-637X/775/1/63}, \href {https://ui.adsabs.harvard.edu/abs/2013ApJ...775...63D} {775, 63}

\bibitem[\protect\citeauthoryear{{Dzib} et~al.,}{{Dzib} et~al.}{2015}]{Dzib+2015}
{Dzib} S.~A.,  et~al., 2015, \mn@doi [\apj] {10.1088/0004-637X/801/2/91}, \href {https://ui.adsabs.harvard.edu/abs/2015ApJ...801...91D} {801, 91}

\bibitem[\protect\citeauthoryear{{Dzib}, {Loinard}, {Ortiz-Le{\'o}n}, {Rodr{\'\i}guez}  \& {Galli}}{{Dzib} et~al.}{2018}]{dzib2018}
{Dzib} S.~A.,  {Loinard} L.,  {Ortiz-Le{\'o}n} G.~N.,  {Rodr{\'\i}guez} L.~F.,   {Galli} P. A.~B.,  2018, \mn@doi [\apj] {10.3847/1538-4357/aae687}, \href {https://ui.adsabs.harvard.edu/abs/2018ApJ...867..151D} {867, 151}

\bibitem[\protect\citeauthoryear{{Esplin}, {Luhman}, {Faherty}, {Mamajek}  \& {Bochanski}}{{Esplin} et~al.}{2017}]{Esplin+2017}
{Esplin} T.~L.,  {Luhman} K.~L.,  {Faherty} J.~K.,  {Mamajek} E.~E.,   {Bochanski} J.~J.,  2017, \mn@doi [\aj] {10.3847/1538-3881/aa74e2}, \href {https://ui.adsabs.harvard.edu/abs/2017AJ....154...46E} {154, 46}

\bibitem[\protect\citeauthoryear{{Feigelson} \& {Lawson}}{{Feigelson} \& {Lawson}}{2004}]{Feigelson+2004}
{Feigelson} E.~D.,  {Lawson} W.~A.,  2004, \mn@doi [\apj] {10.1086/423613}, \href {https://ui.adsabs.harvard.edu/abs/2004ApJ...614..267F} {614, 267}

\bibitem[\protect\citeauthoryear{{Feigelson}, {Casanova}, {Montmerle}  \& {Guibert}}{{Feigelson} et~al.}{1993}]{Feigelson+1993}
{Feigelson} E.~D.,  {Casanova} S.,  {Montmerle} T.,   {Guibert} J.,  1993, \mn@doi [\apj] {10.1086/173264}, \href {https://ui.adsabs.harvard.edu/abs/1993ApJ...416..623F} {416, 623}

\bibitem[\protect\citeauthoryear{{Fomalont}, {Windhorst}, {Kristian}  \& {Kellerman}}{{Fomalont} et~al.}{1991}]{Fomalont+1991}
{Fomalont} E.~B.,  {Windhorst} R.~A.,  {Kristian} J.~A.,   {Kellerman} K.~I.,  1991, \mn@doi [\aj] {10.1086/115952}, \href {https://ui.adsabs.harvard.edu/abs/1991AJ....102.1258F} {102, 1258}

\bibitem[\protect\citeauthoryear{{Frater}, {Brooks}  \& {Whiteoak}}{{Frater} et~al.}{1992}]{Frater+1992}
{Frater} R.~H.,  {Brooks} J.~W.,   {Whiteoak} J.~B.,  1992, Journal of Electrical and Electronics Engineering Australia, \href {https://ui.adsabs.harvard.edu/abs/1992JEEEA..12..103F} {12, 103}

\bibitem[\protect\citeauthoryear{{Gaia Collaboration} et~al.,}{{Gaia Collaboration} et~al.}{2016}]{GaiaDR1}
{Gaia Collaboration} et~al., 2016, \mn@doi [\aap] {10.1051/0004-6361/201629512}, \href {https://ui.adsabs.harvard.edu/abs/2016A&A...595A...2G} {595, A2}

\bibitem[\protect\citeauthoryear{{Gaia Collaboration} et~al.,}{{Gaia Collaboration} et~al.}{2018}]{GaiaDR2}
{Gaia Collaboration} et~al., 2018, \mn@doi [\aap] {10.1051/0004-6361/201833051}, \href {https://ui.adsabs.harvard.edu/abs/2018A&A...616A...1G} {616, A1}

\bibitem[\protect\citeauthoryear{{Galli} et~al.,}{{Galli} et~al.}{2021}]{Galli+2021}
{Galli} P.~A.~B.,  et~al., 2021, \mn@doi [\aap] {10.1051/0004-6361/202039395}, \href {https://ui.adsabs.harvard.edu/abs/2021A&A...646A..46G} {646, A46}

\bibitem[\protect\citeauthoryear{{Graham} \& {Hartigan}}{{Graham} \& {Hartigan}}{1988}]{Graham+1988}
{Graham} J.~A.,  {Hartigan} P.,  1988, \mn@doi [\aj] {10.1086/114715}, \href {https://ui.adsabs.harvard.edu/abs/1988AJ.....95.1197G} {95, 1197}

\bibitem[\protect\citeauthoryear{{Grasdalen}, {Joyce}, {Knacke}, {Strom}  \& {Strom}}{{Grasdalen} et~al.}{1975}]{Grasdalen+1975}
{Grasdalen} G.,  {Joyce} R.,  {Knacke} R.~F.,  {Strom} S.~E.,   {Strom} K.~M.,  1975, \mn@doi [\aj] {10.1086/111720}, \href {https://ui.adsabs.harvard.edu/abs/1975AJ.....80..117G} {80, 117}

\bibitem[\protect\citeauthoryear{{Greisen}}{{Greisen}}{2003}]{AIPS2003}
{Greisen} E.~W.,  2003, in {Heck} A.,  ed.,  Astrophysics and Space Science Library Vol. 285, Information Handling in Astronomy - Historical Vistas. p.~109, \mn@doi{10.1007/0-306-48080-8_7}

\bibitem[\protect\citeauthoryear{{G\"udel} \& {Benz}}{{G\"udel} \& {Benz}}{1993}]{Gudel1993}
{G\"udel} M.,  {Benz} A.~O.,  1993, \mn@doi [\apjl] {10.1086/186766}, \href {https://ui.adsabs.harvard.edu/abs/1993ApJ...405L..63G} {405, L63}

\bibitem[\protect\citeauthoryear{{Hyland}, {Jones}  \& {Mitchell}}{{Hyland} et~al.}{1982}]{Hyland1982}
{Hyland} A.~R.,  {Jones} T.~J.,   {Mitchell} R.~M.,  1982, \mn@doi [\mnras] {10.1093/mnras/201.4.1095}, \href {https://ui.adsabs.harvard.edu/abs/1982MNRAS.201.1095H} {201, 1095}

\bibitem[\protect\citeauthoryear{{Kounkel} et~al.,}{{Kounkel} et~al.}{2014}]{Kounkel+2014}
{Kounkel} M.,  et~al., 2014, \mn@doi [\apj] {10.1088/0004-637X/790/1/49}, \href {https://ui.adsabs.harvard.edu/abs/2014ApJ...790...49K} {790, 49}

\bibitem[\protect\citeauthoryear{{Launhardt}, {Loinard}, {Dzib}, {Forbrich}, {Bower}, {Henning}, {Mioduszewski}  \& {Reffert}}{{Launhardt} et~al.}{2022}]{Launhardt+2022}
{Launhardt} R.,  {Loinard} L.,  {Dzib} S.~A.,  {Forbrich} J.,  {Bower} G.~C.,  {Henning} T.~K.,  {Mioduszewski} A.~J.,   {Reffert} S.,  2022, \mn@doi [\apj] {10.3847/1538-4357/ac5b09}, \href {https://ui.adsabs.harvard.edu/abs/2022ApJ...931...43L} {931, 43}

\bibitem[\protect\citeauthoryear{{Lehtinen}, {Harju}, {Kontinen}  \& {Higdon}}{{Lehtinen} et~al.}{2003}]{Lehtinen+2003}
{Lehtinen} K.,  {Harju} J.,  {Kontinen} S.,   {Higdon} J.~L.,  2003, \mn@doi [\aap] {10.1051/0004-6361:20030185}, \href {https://ui.adsabs.harvard.edu/abs/2003A&A...401.1017L} {401, 1017}

\bibitem[\protect\citeauthoryear{{Luhman}}{{Luhman}}{2008}]{Luhman2008}
{Luhman} K.~L.,  2008, in {Reipurth} B.,  ed., , Vol.~5, Handbook of Star Forming Regions, Volume II.
p.~169, \mn@doi{10.48550/arXiv.0808.3207}

\bibitem[\protect\citeauthoryear{{Mizuno}, {Yamaguchi}, {Tachihara}, {Toyoda}, {Aoyama}, {Yamamoto}, {Onishi}  \& {Fukui}}{{Mizuno} et~al.}{2001}]{Mizuno+2001}
{Mizuno} A.,  {Yamaguchi} R.,  {Tachihara} K.,  {Toyoda} S.,  {Aoyama} H.,  {Yamamoto} H.,  {Onishi} T.,   {Fukui} Y.,  2001, \mn@doi [\pasj] {10.1093/pasj/53.6.1071}, \href {https://ui.adsabs.harvard.edu/abs/2001PASJ...53.1071M} {53, 1071}

\bibitem[\protect\citeauthoryear{{Narang}, {Ohashi}, {Tobin}, {McClure}, {J{\o}rgensen}, {Sai (Insa Choi)}  \& {eDisk + IceAge Team}}{{Narang} et~al.}{2025}]{Narang2025}
{Narang} M.,  {Ohashi} N.,  {Tobin} J.~J.,  {McClure} M.~K.,  {J{\o}rgensen} J.~K.,  {Sai (Insa Choi)} J.,   {eDisk + IceAge Team} 2025, \mn@doi [\aj] {10.3847/1538-3881/adb1ba}, \href {https://ui.adsabs.harvard.edu/abs/2025AJ....169..192N} {169, 192}

\bibitem[\protect\citeauthoryear{{Ord{\'o}{\~n}ez-Toro} et~al.,}{{Ord{\'o}{\~n}ez-Toro} et~al.}{2024}]{Jazmin+2024}
{Ord{\'o}{\~n}ez-Toro} J.,  et~al., 2024, \mn@doi [\aj] {10.3847/1538-3881/ad1bd3}, \href {https://ui.adsabs.harvard.edu/abs/2024AJ....167..108O} {167, 108}

\bibitem[\protect\citeauthoryear{{Ortiz-Le{\'o}n} et~al.,}{{Ortiz-Le{\'o}n} et~al.}{2015}]{OrtizLeon+2015}
{Ortiz-Le{\'o}n} G.~N.,  et~al., 2015, \mn@doi [\apj] {10.1088/0004-637X/805/1/9}, \href {https://ui.adsabs.harvard.edu/abs/2015ApJ...805....9O} {805, 9}

\bibitem[\protect\citeauthoryear{{Ortiz-Le{\'o}n} et~al.,}{{Ortiz-Le{\'o}n} et~al.}{2017}]{OrtizLeon+2017}
{Ortiz-Le{\'o}n} G.~N.,  et~al., 2017, \mn@doi [\apj] {10.3847/1538-4357/834/2/141}, \href {https://ui.adsabs.harvard.edu/abs/2017ApJ...834..141O} {834, 141}

\bibitem[\protect\citeauthoryear{{Pech} et~al.,}{{Pech} et~al.}{2016}]{Pech+2016}
{Pech} G.,  et~al., 2016, \mn@doi [\apj] {10.3847/0004-637X/818/2/116}, \href {https://ui.adsabs.harvard.edu/abs/2016ApJ...818..116P} {818, 116}

\bibitem[\protect\citeauthoryear{{Reid} et~al.,}{{Reid} et~al.}{2017}]{Reid+2017}
{Reid} M.~J.,  et~al., 2017, \mn@doi [\aj] {10.3847/1538-3881/aa7850}, \href {https://ui.adsabs.harvard.edu/abs/2017AJ....154...63R} {154, 63}

\bibitem[\protect\citeauthoryear{{Robrade} \& {Schmitt}}{{Robrade} \& {Schmitt}}{2007}]{Robrade+2007}
{Robrade} J.,  {Schmitt} J.~H.~M.~M.,  2007, \mn@doi [\aap] {10.1051/0004-6361:20066250}, \href {https://ui.adsabs.harvard.edu/abs/2007A&A...461..669R} {461, 669}

\bibitem[\protect\citeauthoryear{{Roccatagliata}, {Sacco}, {Franciosini}  \& {Randich}}{{Roccatagliata} et~al.}{2018}]{Rocca+2018}
{Roccatagliata} V.,  {Sacco} G.~G.,  {Franciosini} E.,   {Randich} S.,  2018, \mn@doi [\aap] {10.1051/0004-6361/201833890}, \href {https://ui.adsabs.harvard.edu/abs/2018A&A...617L...4R} {617, L4}

\bibitem[\protect\citeauthoryear{{Sault}, {Teuben}  \& {Wright}}{{Sault} et~al.}{1995}]{MIRIAD1995}
{Sault} R.~J.,  {Teuben} P.~J.,   {Wright} M.~C.~H.,  1995, in {Shaw} R.~A.,  {Payne} H.~E.,   {Hayes} J.~J.~E.,  eds,  Astronomical Society of the Pacific Conference Series Vol. 77, Astronomical Data Analysis Software and Systems IV. p.~433 (\mn@eprint {arXiv} {astro-ph/0612759}), \mn@doi{10.48550/arXiv.astro-ph/0612759}

\bibitem[\protect\citeauthoryear{{Schwartz}}{{Schwartz}}{1977}]{Schwartz1977}
{Schwartz} R.~D.,  1977, \mn@doi [\apjs] {10.1086/190473}, \href {https://ui.adsabs.harvard.edu/abs/1977ApJS...35..161S} {35, 161}

\bibitem[\protect\citeauthoryear{{Schwartz}}{{Schwartz}}{1992}]{Schwartz1992}
{Schwartz} R.~D.,  1992, in {Reipurth} B.,  ed., , Low Mass Star Formation in Southern Molecular Clouds.
p.~93

\bibitem[\protect\citeauthoryear{{Spezzi} et~al.,}{{Spezzi} et~al.}{2008}]{Spezzi+2008}
{Spezzi} L.,  et~al., 2008, \mn@doi [\apj] {10.1086/587931}, \href {https://ui.adsabs.harvard.edu/abs/2008ApJ...680.1295S} {680, 1295}

\bibitem[\protect\citeauthoryear{{Stelzer}, {Micela}  \& {Neuh{\"a}user}}{{Stelzer} et~al.}{2004}]{Stelzer+2004}
{Stelzer} B.,  {Micela} G.,   {Neuh{\"a}user} R.,  2004, \mn@doi [\aap] {10.1051/0004-6361:20040202}, \href {https://ui.adsabs.harvard.edu/abs/2004A&A...423.1029S} {423, 1029}

\bibitem[\protect\citeauthoryear{{Thompson}, {Moran}  \& {Swenson}}{{Thompson} et~al.}{2017}]{Moran+2017}
{Thompson} A.~R.,  {Moran} J.~M.,   {Swenson} Jr. G.~W.,  2017, {Interferometry and Synthesis in Radio Astronomy, 3rd Edition}, \mn@doi{10.1007/978-3-319-44431-4.
}

\bibitem[\protect\citeauthoryear{{Voirin}, {Manara}  \& {Prusti}}{{Voirin} et~al.}{2018}]{Voirin+2018}
{Voirin} J.,  {Manara} C.~F.,   {Prusti} T.,  2018, \mn@doi [\aap] {10.1051/0004-6361/201731153}, \href {https://ui.adsabs.harvard.edu/abs/2018A&A...610A..64V} {610, A64}

\bibitem[\protect\citeauthoryear{{Whittet}, {Prusti}, {Franco}, {Gerakines}, {Kilkenny}, {Larson}  \& {Wesselius}}{{Whittet} et~al.}{1997}]{Whittet+1997}
{Whittet} D.~C.~B.,  {Prusti} T.,  {Franco} G.~A.~P.,  {Gerakines} P.~A.,  {Kilkenny} D.,  {Larson} K.~A.,   {Wesselius} P.~R.,  1997, \aap, \href {https://ui.adsabs.harvard.edu/abs/1997A&A...327.1194W} {327, 1194}

\bibitem[\protect\citeauthoryear{{Wilson} et~al.,}{{Wilson} et~al.}{2011}]{CABB2011}
{Wilson} W.~E.,  et~al., 2011, \mn@doi [\mnras] {10.1111/j.1365-2966.2011.19054.x}, \href {https://ui.adsabs.harvard.edu/abs/2011MNRAS.416..832W} {416, 832}

\bibitem[\protect\citeauthoryear{{Yiu}, {Vedantham}, {Callingham}  \& {G{\"u}nther}}{{Yiu} et~al.}{2024}]{Yiu+2024}
{Yiu} T.~W.~H.,  {Vedantham} H.~K.,  {Callingham} J.~R.,   {G{\"u}nther} M.~N.,  2024, \mn@doi [\aap] {10.1051/0004-6361/202347657}, \href {https://ui.adsabs.harvard.edu/abs/2024A&A...684A...3Y} {684, A3}

\makeatother
\end{thebibliography}
